\documentclass{elsart}

\usepackage{natbib}

 \usepackage{graphicx}
\usepackage{amssymb}
\usepackage{amsmath}
\newcommand{\fr}{\frac}

\begin{document}

\begin{frontmatter}


\title{On the tilt of Fundamental Plane by Clausius' virial maximum theory}

%\address[label1]{*}

\author{L. Secco*, D. Bindoni}
%\address[io]{Dipartimento di Astronomia, Universit\`a di Padova, Padova, Italy}
\address{Department of Astronomy, University of Padova, Padova, Italy}
\ead{luigi.secco@unipd.it}

\begin{abstract}

%\begin{abstract}
% Text of abstract
The theory of the Clausius' virial maximum to explain the Fundamental Plane (FP) proposed by  
Secco (2000, 2001,2005) is based on the existence of a 
maximum in the Clausius' Virial (CV) potential 
energy of a early type galaxy (ETG) stellar component when it is 
completely embedded inside a dark matter (DM) halo. At the first order approximation the 
theory was developed by modeling the two-components with two cored power-law density 
profiles.
An higher level of approximation is now taken into account by developing the same 
theory when the stellar component is modeled by a King-model with a cut-off. 
Even if the DM halo density remains a cored power law the inner component is now more
realistic for the ETGs. The new formulation allows us to understand more deeply what is the 
dynamical reason of the FP {\it tilt}  and in general 
how the CV theory may really be the engine  to produce the FP main features. The degeneracy of FP
in respect to the initial density perturbation spectrum may be now full understood in
a CDM cosmological scenario. 
A possible way to compare the FPs predicted by the theory with those obtained by observations is also 
exemplified.

\end{abstract}


\end{frontmatter}
\section{On the tilt}

It is well known that galaxies of different morphological types cluster 
around the Fundamental Plane (FP) (Dressler et al., 1987; Djorgovski and Davies, 1987;
Faber et al., 1987; Bender et al., 1992; Djorgovski \& Santiago, 1993;
Renzini \& Ciotti, 1993; Ciotti et al.
1996; J{\o}rgensen, 1999; see, e.g., the review of D'Onofrio et al., 2006, and the 
references therein) in the three dimensional space of: $r_{e}$, effective 
radius; $I_{e}$, the mean effective surface brigtness within $r_{e}$; $\sigma_{o}$, the 
central projected velocity dispersion. On the basis of \textit{homology + virial theorem} 
one would expect that the FP equation has to be: $r_{e}\sim \sigma^{A}_{o} I^{B}_{e}$ 
where $A=2$, $B=-1$. That results completely in disagreement with the observations in 
different bands. Typical values in $B-$band are: $A=1.33\pm 0.05$; $B=-0.83\pm 0.03$
(e.g., in D'Onofrio et al., 2006).
These unexpected values produce in the $\kappa$ coordinate system (Bender et al.,1992) the so 
called \textit{tilt} that is an increasing of the ratio: dynamical mass $M_{dyn}$ over 
luminosity $L$, of this kind:
\begin{equation}
\label{tilt}
M_{dyn}/L\sim (M_{dyn})^{0.2}
\end{equation}

Many attempts have been done in order to understand the FP {\it tilt} which is also 
one of the common features either for galaxy FPs or 
for the FPs of all virialized structures  which all together define the so called 
{\em cosmic meta-plane} (Burstein et al., 1997).
The review of D'Onofrio et al. (2006) may help the reader to take into account 
the more recent efforts to solve the hard problem of finding an explanation of the trend (\ref{tilt}) 
when the $K-band$ is also considered and then the population effect has to be rouled out.
Actually 
it is possible to explain the trend observed in the $B-$band as a metallicity 
sequence of an old stellar population (Maraston, 1999). However the $M_{dyn}/L$ 
values in the $K-$band are independent of metallicity even if the tilt 
is observed (Pahre et al., 1998). A secondary effect is then needed to explain the $K-$band tilt (Gerhard et al., 2001).

The Clausius' Virial theory (TCV) of FP has the aim to propose 
a dynamical mechanism able to produce the required effect on a huge range of 
mass scales from globular clusters
to galaxy clusters.
 The purpose is to prove that it may be possible to change $A,B$ exponents 
 (from the expected values $2,-1$) without breaking \textit{homology + virial 
 equilibrium}. It is based on the existence of a special virial configuration characterized
 by a maximum in the Clausius' Virial 
 potential energy (CV) which, on galaxy mass scale, refers to a 
 baryonic (stellar, $B$) component when it is completely embedded inside a DM halo ($D$ component).
At the first order approximation (linear) the two-components are modelled with two 
power-law density profiles and two infinitesimal cores.
The general strategy is 
described in many papers 
(Secco, 2000; Secco, 2001, hereafter LS1; Marmo \& Secco, 2003; Secco, 2005, hereafter LS5).

Now we move from a linear approach of TCV to a non-linear one
that is to an higher level of approximation in which 
the stellar component is built up by a King-model with a cut-off. 
Even if the DM halo density remains a power law the inner component is actually more
realistic for the ETGs. The new formulation allows us to understand more deeply the physical reasons
which produce the FP {\it tilt} and the role of the main involved quantities, particularly that of
$I_e$. Moreover we may begin the comparison 
between the expected edge-on  FPs with those obtained by observations (e.g. that of Djorgovski \&
Davies (1987)) and try to reproduce in $\kappa$-space the {\it tilt} fit-equation of
Burstein et al. (1997). Its theoretical derivation may explain why the FP is degenerate 
in respect to the initial density perturbation spectrum in a CDM scenario as already underlined by Djorgovski (1992).
Some initial examples of theoretical FP calibration will
be given in the sects. 7, 9, for some special choice of theoretical parameters.
A more complete discussion is still in progress. 

\section{General strategy of TCV}
Briefly summarizing,
the general strategy consists to use the two-component tensor virial theorem 
(e.g., Brosche et al., 1983; Caimmi \& Secco, 1992) to describe the virial configuration
of the baryonic component embedded in a DM halo at the end of relaxation phase 
(see, Bindoni \& Secco, 2008).
It reads:
\begin{equation}
\label{vireq} 
2(T_{u})_{ij}=(V_{u})_{ij}; (u=B,D; i,j= x,y,z)
\end{equation}
According to the scalar virial for one component, the potential energy tensor, which has to enter into
the tensor virial equations, is the Clausius' virial tensor, $(V_{u})_{ij}$, built-up of the \textit{self potential-energy tensor}, $(\Omega_u)_{ij}$, and the \textit{tidal potential-energy tensor}, $(V_{uv})_{ij}$. 
Then, according to the scalar virial theorem, the trace of CV tensor, in the case of stellar component, has to be read: 

\begin{eqnarray}\
\label{Clau}
V_B= \Omega_B + V_{BD}\\
\label{omB}
\nonumber
\Omega_B= \int\rho_B\sum_{r={1}}^{3}~ x_r\frac{\partial\Phi_B}
{\partial x_r}{\rm\,d}\vec{x_B}~= \int\rho_B
 (\vec{r_B}\cdot\vec{f_B}){\rm\,d}\vec{x_B}\\
\label{VBA}
\nonumber
(V_{BD})= \int\rho_B\sum_{r={1}}^{3} x_r\frac{\partial\Phi_D}
{\partial x_r}{\rm\,d}\vec{x_B}~=\int\rho_B(\vec{r_B}\cdot\vec{f_D})
{\rm\,d}\vec{x_B};
\end{eqnarray}

where $\rho_{B}$ is the $B$ component density and $\vec{f}_{B}, \vec{f}_{D}$ are the force per unit mass due to the 
self and DM gravity, respectively, at the point $\vec{r}_B$ and  $\Phi_B$, $\Phi_D$ are the respective potentials.

Conversely, the total potential energy tensor of the $B$ component is:
$(\Omega_B)_{ij} + (W_{BD})_{ij}$,
where the interaction energy tensor is:
$(W_{BD})_{ij}=-\frac12\int\rho_{B}(\Phi_{D})_{ij}{\rm\,d}\vec{x}_{B}$; and
the potential tensor due to the DM (e.g., Chandrasekhar, 1969) is:
$(\Phi_{D})_{ij}=G \int\rho_{D}(\vec{x^{\prime}})\frac
{(x_{i}-x^{\prime}_{i})( x_{j}-x^{\prime}_{j})}{\mid {\vec{x}-\vec{x^{\prime}}\mid}^3}
{\rm\,d}\vec{x}_{D}$.

To be noted that in general: $(V_{BD})_{ij}\neq(W_{BD})_{ij}$, the difference gives the residual energy tensor (Caimmi \& Secco, 1992).

We will describe a re-formulation of TCV   
in the case in which the two-component model is built up of: a bright $B$ stellar 
component with a King (1962) truncated density profile completely embedded in a 
DM frozen halo, $D$, with a cored power-law mass density distribution. 

\section{Why introducing King's model}
\subsection{End of relaxation phase}

The violent relaxation mechanism leads 
to an equipartition of energy per unit mass and not per particles (see, e.g., the review of
Bindoni \& Secco, 2008, and references therein).
 If $\sigma $ is the velocity dispersion, assumed to be
the same for every star mass, integration of the distribution function, $f(E)$, over the velocities
(Binney \& Tremaine, 1987, Chapter 4; Combes, 1995, Chapter 4), yields the density:
\begin{equation}
\label{rho}
\rho(r)=\rho_1 e^{-U(r)/\sigma^2}
\end{equation}
where the total energy per unit mass is: $E=(1/2)v^2 +U$; ($v$ and $U$ are velocity and
potential 
energy per unit mass, respectively).
On the other hand, Poisson equation:

\begin{equation}
\label{pois}
\frac{1}{r^2}\frac{d}{dr}(r^2 \frac{dU}{dr})=4 \pi G \int  f(E)d\vec{v}
\end{equation}
becomes by means of Eq.(\ref{rho}):
\begin{equation}
\label{binn}
\frac{d}{dr}(r^2 \frac{d\ln\rho}{dr})=-\frac{4 \pi G }{\sigma^2} r^2\rho 
\end{equation}
with the solution:
\begin{equation}
\label{sol}
\rho(r)=\frac{\sigma^2}{2\pi Gr^2}
\end{equation}
In turn, Eq. (\ref{rho}) gives:
\begin{equation}
\label{corera}
2\ln(\frac{3}{\sqrt{2}}\frac{r}{r_c})=U(r)/ \sigma^2 
\end{equation}
when a core radius $r_c= 3\sigma(4 \pi G \rho_o)^{-1/2}$ is introduced in order to avoid 
an infinite value of the central density $\rho_o$. $r_c$ corresponds to the radius at 
which the projected 
density of the isothermal sphere falls to roughly half of its central value.
Eq.(\ref{corera}) gives us the asymptotic behavior as soon as $r$ is greater of about $2r_c$:
\begin{equation}
U(r)\approx 2\sigma^2\ln(r/r_c)
\end{equation}
which means again from Eq.(\ref{rho}), an isothermal behavior, $\rho(r)\propto r^{-2}$ as $r\rightarrow\infty$.

\subsection{Problems with isothermals}
The isothermal energy distribution function extends spatially to infinity with infinite mass
and so
does not be suitable to represent a real elliptical galaxy.

Since 1965 
Ogorodnikov has highlighted that: in order to find the most probable phase distribution function
for a stellar system in a stationary state, the phase volume has to be truncated  in both
coordinate and velocity space. While in the velocity space the truncation 
arises spontaneously due to the existence of escape velocity, 
the introduction of a cut-off in the 
coordinate space appears, on one side,
necessary in order to obtain a finite mass $M$ and radius $R$, but, on the
other, very problematic. 

A similar 
difficulty also 
appears on the
thermodynamical side, for which an extensive literature exists (from: Lynden-Bell
\& Wood, 1968; Horowitz \& Katz, 1978; White \& Narayan, 1987, until, e.g., 
Bertin \& Trenti, 2003, and references therein). By using 
the standard Boltzmann-Gibbs entropy:
\begin{equation}
\label{entr}
S=-\int {f\ln f d^3x d^3v}
\end{equation}
defined by
the {\it distribution function}, $f(\vec{x},\vec{v})$ (hereafter $DF$), in the $\mu$ 
phase-space, 
and looking for
what maximizes the entropy of the same stellar system, the conclusion is: the $DF$
which plays this role in (\ref{entr}) is that of the isothermal sphere. But, the 
maximization of ${\bf S}$, subject to fixed mass $M$ and energy $E$, leads again to a $DF$ that is incompatible with
finite $M$ and $E$ (see, e.g., Binney
\& Tremaine, 1987, Chapter 4; Merritt 1999, Lima Neto et al. 1999, Marquez et al. 2001, 
and references therein).

Our limited contribution to the wide discussion existing in the literature will be
to underline as in a 
stellar component, 
embedded 
in a second dark 
matter subsystem (e.g., Ciotti, 1999, and references therein), 
a truncation is spontaneously introduced in coordinate space, due to the presence of 
a scale length induced
from the dark halo, as long as virial equilibrium holds.
That is the {\it tidal radius} which has been discovered in the TCV dynamical theory
(LS1, LS5) we will revisite in the next paragraphs.\footnote{
Even if some considerations which follow are more general and may also be
extended to spirals, we will limit our considerations 
to the collisionless stellar systems, as the ellipticals are considered.}

\subsection{King's models with cut-off}
We will assume for the B-component the empirical surface light density law proposed by King (1962) as:
\begin{equation}
\label{king}
I(R)=k_L\left\{\frac{1}{[1+(R/R_c)^2]^{1/2}}-\frac{1}{[1+(R_t/R_c)^2]^{1/2}}\right\}^{2}
\end{equation}
where $R_t$ is the value of $R$ at which $I$ reaches zero. The law has the advantage to take 
into account the existence of a cut-off in the surface density 
distribution as one expects for globular clusters (GC) (the profile (\ref{king}) is born 
for them) but also
for ellipticals too. Indeed we will need of a similar truncation radius because in the 
TCV theory the meaning of the radius at the special Clausius virial maximum
configuration has an analogous role 
(LS5) of that discovered by von Hoerner (1958) for GC. The last one is due to the tidal effect of 
Galaxy, the former one  to a similar 
tidal effect but, this time, due to the dynamical effect of DM halo
distribution on the baryonic component. The value $R_c$ corresponds to the 
core-radius and the $k_L$ value is linked to the central surface light density $I_o$ by:
\begin{equation}
\label{fo}
I_o=k_L\left\{1-\frac{1}{[1+(R_t/R_c)^2]^{1/2}}\right\}^{2}
\end{equation}

\section{King's model in phase-space}
From the previous considerations,
an acceptable distribution function in the phase-space has to have a cut-off at the energy $E_o$ as King introduced in 
his model (1966):
\begin{eqnarray}
f_K(E)=0 \ for \ E \ge E_o\\
f_K(E)= (2\pi\sigma^2)^{-3/2}\rho_o(e^{(E_o-E)/\sigma^2}-1) \ for \ E < E_o
\end{eqnarray}
The spatial density can be obtained after integration over the velocities
in the following way:
\begin{equation}
\frac{\rho(r)}{\rho_o}=e^y \text{erf}(y^{1/2})-\left (\frac{4y}{\pi}\right)^{1/2}
\left(1+\frac{2y}{3}\right);\ y=-U/\sigma^2
\end{equation}
\text{erf} being the error function, $$\text{erf}(x)=\frac{2}{\sqrt{\pi}}\int_{0}^{x}e^{-u^2}du$$

Following King (1962), the spatial density is given by:
\begin{eqnarray}
\label{masD}
	\rho_{B}\left(z\right)=\frac{k_{M}}{\pi r_{c}}\frac{z_o^3}{z^{2}}\left[\frac{1}{z}\cos^{-1}z-\left(1-z\right)^{\frac{1}{2}}\right]
\end{eqnarray}
where
\begin{eqnarray}
\label{zed}
z=\left[\frac{1+\left(r/r_{c}\right)^{2}}{1+\left(r_{t}/r_{c}\right)^{2}}\right]^{\frac{1}{2}}; z_o=\left[\frac{1}{1+\left(r_{t}/r_{c}\right)^{2}}\right]^{\frac{1}{2}}
\end{eqnarray}
with $r_c$ (=$R_c$) and $r_t$ (=$R_t$) are  the core and the cut-off spatial radius, respectively. The mass inside $z$ is given by:
\begin{equation}
\label{fmas}
M(z)=4\pi r_c^2 k_M z_o\int_{z_o}^{z}\left[\frac{z^2}{z_o^2}-1\right]\frac{1}{z}\left[cos^{-1}z -(1-z^2)^{1/2}\right]dz
\end{equation} 

The projected density is:

\begin{equation}
\label{masP}
\Sigma (R)=k_{M}\left(\frac{1}{[1+(\frac{R}{R_c})^2]^{1/2}}
-\frac{1}{[1+(\frac{R_t}{R_c})^2]^{1/2}}\right)^2
\end{equation}
which is linked to the spatial density by the Abel integral equation (Binney \& Tremaine, 1987, Chapt.4):
\begin{equation}
\label{abel}
\rho(r)=-\frac{1}{\pi}\int_r^{r_t}\frac{d\Sigma}{dR}\frac{dR}{\sqrt{R^2-r^2}}
\end{equation}

\begin{figure}[!ht]
\begin{center}
\includegraphics[width=1.00\textwidth]{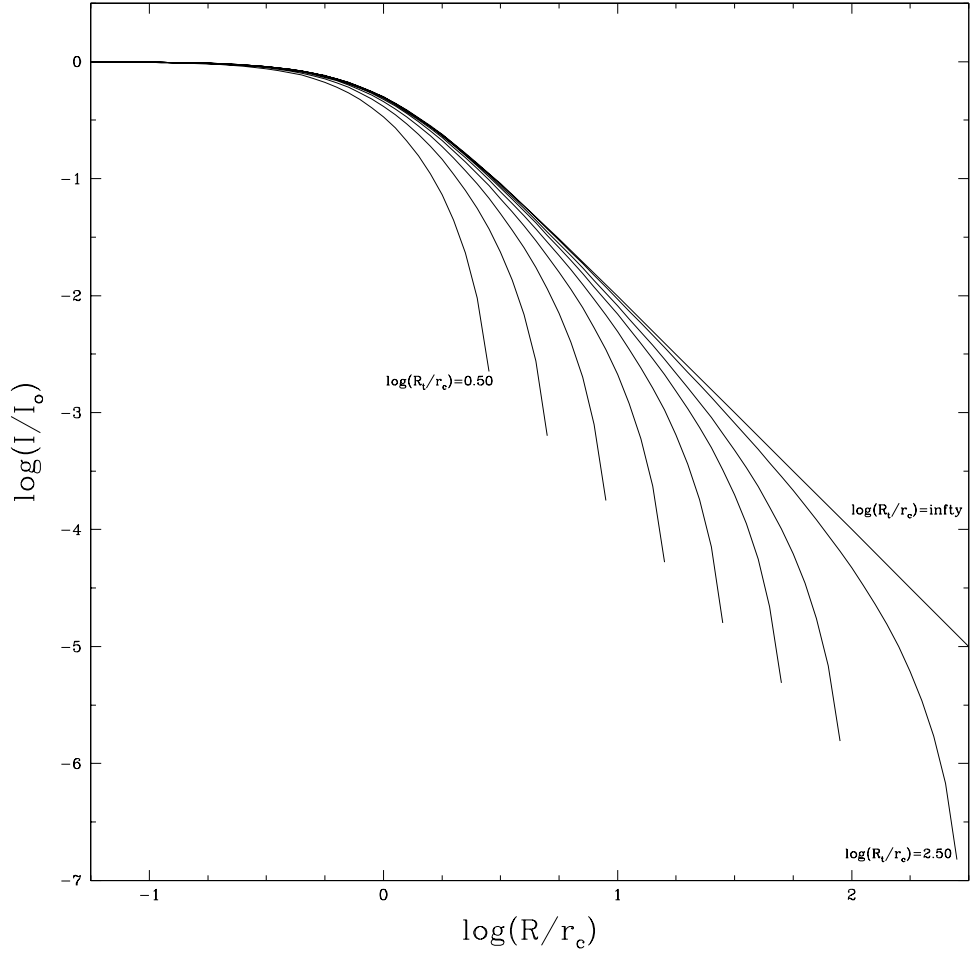}
\end{center}

\caption{The standard curves of King's model of $I(R)$ normalized to the central value as function of the different
parameters $R_t/r_c$. The limit case when the cut-off goes to $\infty$ is also shown (King, 1962).} 
\label{fig:kingprofile}

\label{fig01}
\end{figure}

$\Sigma (R)$ becomes the classical surface luminosity density $I(R)$ given by King (1962) 
(Eq.\ref{king})
 as soon as $k_M$ translates into
$k_L$ (see, Fig.\ref{fig:kingprofile}). It should be noted that 
in our models $k_M \ne k_L$ being one of our main
assumptions. Indeed in the TCV
the galaxy homology family is intrinsically  characterized by a ratio mass/luminosity for the 
$B$ component
different from a constant.

Integration of $\Sigma (R)$  with respect of $2\pi R dR$ gives the total
projected mass within the projected distance $R$ of the center which becomes the 
luminosity function, $L(X)$, with the substitution of $\Sigma (R)\rightarrow I(R)$ and $k_M\rightarrow k_L$:

\begin{equation}
\label{limL}
	L\left(X\right)=\pi r^{2}_{c}k_{L}F_L(X)
\end{equation}
where:

\begin{eqnarray}
\label{flx}
F_L(X)=\left[\ln\left(1+X\right)-
	4\frac{\left(1+X\right)^{1/2}-1}{\left(1+X_{t}\right)^{1/2}}+\frac{X}{1+X_{t}}\right]\\
X=\left(\frac{R}{r_{c}}\right)^{2}\ \ \ ,\ \ \ X_{t}=\left(\frac{R_{t}}{r_{c}}\right)^{2}
\end{eqnarray}

According to Eq.(\ref{limL}), as $X_t>>1$ the limit of $L(X_t)$ goes approximately to:
\begin{equation}
\label{LL}
	L\left(X_t\right)\simeq\pi r^{2}_{c}k_{L}\ \ln\left(\frac{r_t^2}{20 r_c^2}\right)
\end{equation}

It should be noted that, according to Eqs.(\ref{fmas}, \ref{limL}), the following equation has to hold:

\begin{equation}
\label{m/l}
\frac{M(1)}{r^2_c k_M}k_L=\frac{L(X_t)}{r_c^2}\rightarrow M_B/L=k_M/k_L
\end{equation}

\begin{figure}[!ht]
\begin{center}
\includegraphics[width=1.00\textwidth]{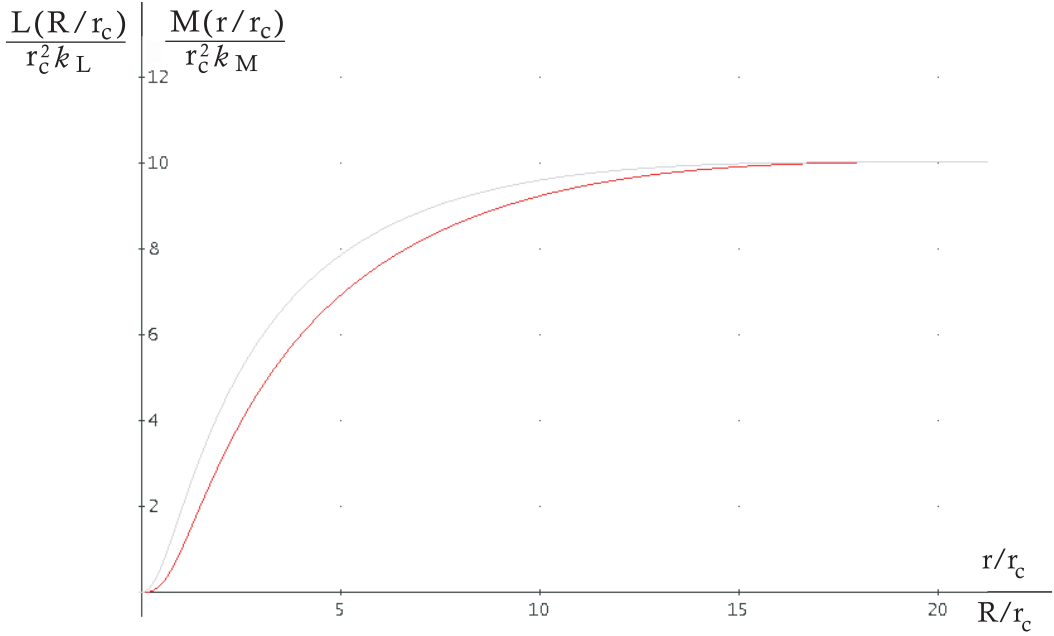}
\end{center}

\caption{Trends of $\frac{L(R/r_c)}{r^2_ck_L}$ and $\frac{M(r/r_c)}{r^2_c k_M}$ in 
the case of, $r_t/r_c=20$,
by assuming $k_M=k_L$. The upper curve is corresponding to the integration of surface luminosity
density and the bottom one to the spatial mass density integration. At $R_t=r_t=20 r_c$ 
the Eq.(\ref{m/l}) holds.}

\label{fig2}
\end{figure}

The trends of normalized $L(R/r_c)$ and $M(r/r_c)$ are shown in the Fig.(\ref{fig2}), in the case of $r_t/r_c=20$,
by assuming $k_M=k_L$.

\section{Inserting King's model into TCV}

The aim is to translate the linear formulation of TCV into
a non-linear one modeling the bright (or baryonic) inner component, $B$, of mass $M_B$ with
a King's model and the dark matter halo, $D$, of mass $M_D$ with a cored power-law.

The first step is to build up the Clausius'virial tensor trace of Eq.(\ref{Clau})
which 
for similar strata spheroids is given by:
\begin{equation}
\label{vir}
V_B=-\nu_{\Omega B}\frac{GM_B^2}{a_B}F-\nu_V\frac{GM_B^2}{a_B}F
\end{equation}
where $\nu_{\Omega B}$ and,
$\nu_V$, are the self mass distribution and 
the interaction coefficient, respectively.
$F$ is a form factor which is equal $2$ in the spherical case we choose for the sake 
of simplicity.

We will follow the general method proposed by Caimmi (1993).
Then we  
define (Roberts, 1962; Caimmi \& Marmo, 2003, and references therein):
$$F_u(\xi_u)=2\int_{\xi_u}^{1}f_u(\xi_u)\xi_ud\xi_u~~;
u=B,D~~$$
where
$f_u(\xi_u)$ are the dimensionless density profiles with:
$$\xi_u= r/a_u\rightarrow C_u\xi_u=
r/r_{ou}\rightarrow C_u=a_u/r_{ou}$$
$C_u,a_u, r_{ou}$ being the concentration, virial and scale radius, respectively. For the King's
inner component 
$a_B$ is the cutoff radius $r_t$, $r_{oB}=r_c$, $C_B=r_t/r_c$.

The King's profile, normalized to the scale radius density value, omiting for sake of simplicity
some
obvious indices, is: 

\begin{eqnarray}
f_B(\xi)
	=\frac{2}{1+(C \xi )^2}\cdot\frac{1}{H}~~~~~~~~~~~~~~~~~~~~~~~~~~~~~~~~~~~~~~~~\\
\nonumber	
	\left
	(\left[\frac{1+C^2}{1+(C \xi )^2}\right]^{1/2}
	\cdot
	\cos^{-1}
	\left[\left(\frac{1+(C \xi )^2}{1+C^2}\right)^{1/2}\right]
	-\left[\frac{
	C^2-(C \xi )^2}{1+C^2}\right]^{1/2}\right );~~~~~~~\\
H=\left\{\left[\frac{1+C^2}{2}\right]^{\frac{1}{2}}\cdot\cos^{-1}
\left[\left(\frac{2}{1+C^2}\right)^{\frac{1}{2}}\right]-
\left[\frac{C^2-1}{1+C^2}\right]^{\frac{1}{2}}\right\}~~~~~~
\end{eqnarray}

The power-law profile for DM, normalized again to $\rho_o(\xi=1/C_D)$, becomes:
\begin{equation}
\label{dmp}
f_D(\xi_D)=\frac{2}{1+(C_D\xi_D)^{d}}
\end{equation}

Now we can define all the coefficients we need for the 
Clausius trace:

\begin{equation}
\label{CoeIn}
\nu_{V}(x)=-\frac98\frac1{(\nu_B)_{M}(\nu_D)_{M}}m
w^{(ext)}(x)~~;x=a_B/a_D
\end{equation}

\begin{equation}
\label{nu}
(\nu_u)_{M}=\frac32\int_0^{1}F_u(\xi_u)d\xi_u~~;
u=B,D~~; 
\end{equation}
and:
\begin{equation}
\label{cnum}
\nu_{\Omega u}=\frac{9}{16}(\nu_u)_M^{-2}\int_0^{1}F_u^2(\xi_u)d\xi_u~~
\end{equation}
The mass ratio Dark/Bright is:
\begin{equation}
m=\frac{M_D}{M_B}
\end{equation}
and the interaction term inside the tidal tensor 
trace, $V_{BD}$, (via the coefficient $\nu_V$ of Eq.(\ref{CoeIn})) due to
dynamical effect of $DM$ on the baryonic one, is given by:
\begin{equation}
\label{Coe}
w^{(ext)}(x)=\int_0^{x}F_B
(\xi_B)\frac{d F_D}{d\xi_D}\xi_D d\xi_D~~;\xi_B=\xi_D/x~ 
\end{equation}

where $(\nu_u)_{M}$ is an additional profile factor which gives the mass of
the two components:
\begin{equation}
\label{Zmass}
M_u=(\nu_u)_{M}M_{ou};~~~ M_{ou}=\frac{4\pi}{3}\rho_{ou} a_u^3
\end{equation}
The main functions cited here are explicitely given in Appendix.

\section{Main features of TCV revisited}
Before to draw the main lines of TCV let us to depict the cosmological
environment inside which the approach, based on tensor virial theorem extended
to two components, tries to explain the galaxy FP-{\it tilt}. As we have showed in LS1, it 
is impossible to give account of the galaxy scaling relationships without
considering the cosmic scenario even if 
the FP as a whole appears to contain a degeneracy in respect to the initial density
perturbation spectrum.That has been pointed out first by Djorgovski (1992).
We try to explain why in the sect. 9.
\subsection{Cosmological framework}

Our framework is a hierarchical CDM scenario (see, e.g., Coles \& Lucchin, 1995).
We assume that 
the spherically-averaged properties of a galaxy assembled dark halo of mass $M_D$
formed by
hierarchical clustering, may be deduced from the linear theory where the mass variance, $\sigma _{M_D}$, in a random Gaussian field
of an Einstein-de Sitter model (Silk, 1999, Chap.3), evolves from the 
recombination time $t_{rec}$ forwards, as :
\begin{equation}
\label{lcosm}
\sigma_{M_D}(t)\sim M_D^{-(n+3)/6} (t/t_{rec})^{2/3}
\end{equation}
where $n=n_{rec}$ is the effective index. This is a self-similar toy model in which,
if the linear regime ends at maximum expansion time, $t_{max}$ (i.e., the free-fall time $\tau_{ff}$)
of the spherical top-hat filtered mini-universe of comoving radius $a_D$,
mass $M_D$ and density $\rho_D$, it holds: $\rho_{Do}^{-1/2}\sim \tau_{ff}\sim (M_D/a_{Do}^3)^{-1/2}$, where $a_D=a_{Do}$
and $\rho_D=\rho_{Do}$ at $t_{max}$.
From Eq.(\ref{lcosm}) it follows that:
\begin{equation}
\tau_{ff}\sim \sigma_{M_D}(t_{rec})^{-3/2}\sim M_D^{\frac{3}{2}\alpha}
\end{equation}
where the local slope:
\begin{equation}
\alpha=\alpha_{rec}=-\frac{d\ln\sigma_{M_D}(t_{rec})}{d\ln M_D}=(n_{rec}+3)/6
\end{equation}
 and then, in turn:
 \begin{equation}
a_{Do}\sim M^{1/\gamma'},\  1/\gamma'=(5+n_{rec})/6=\frac{3\alpha +1}{3}
\end{equation}

The "formation" of halo, i.e., its virialization occurs at
$\delta \rho/\rho=1.67$ when the time from $t_{rec}$ is about $t_F=2\tau_{ff}$ and its radius
becomes:
\begin{equation}
\label{adM}
a_D\simeq a_{Do}/2\sim M^{1/\gamma'}
\end{equation}
That is because  the collisionless halo system relaxes under the violent
relaxation mechanism by conserving its total energy. This is also true even if baryons and dark matter particles 
relaxe together at least under some conditions (see, e.g., LS1, parag.8). Moreover, at $t_F$ the mean density
$\bar{\rho}_F$ is approximately $180\rho_u$ where $\rho_u=1/(6\pi G t_F^2)$. Due to the scaling of
$t_F\sim M_D^{(n+3)/4}$ proved before, at virial equilibrium it holds: $\bar{\rho}_F\sim M_D^{-(n+3)/2}$.
As soon as the halo density profile is of Zhao (1996) family, from Eq.(\ref{Zmass}) the mean density is given
by $(\nu_D)_M\rho_{oD}$ and then if the concentration $C_D$, on which $(\nu_D)_M$ depends, correlates 
loosely with the mass (Dolag et al., 2004), the central density
has to scale as: 
\begin{equation}
\rho_{oD}\sim M_D^{-(n+3)/2}
\end{equation}
 Then low mass halos are significantly denser
than more massive systems. That reflects the higher collapse redshift of small halos (Navarro et al., 1997). 
Indeed the relationship between mass and formation redshift, $z_F$, defined as the $z$ at which an object of present mass $M_D$ has on average acquired half its mass
is (Lacey \& Cole, 1993):
\begin{equation}
\label{redsh}
z_F= (2^{(n+3)/3}-1)^{1/2}(M_D/M_{nl})^{-(n+3)/6}  
\end{equation}
$M_{nl}= 4\cdot10^{13}(1+z)^{-6/(3+n)} M_{\odot}$, is the
mass scale over which galaxy count fluctuations have unit variance, 
corresponding to the comoving, $R_{nl}=8 h^{-1} Mpc$.
These scaling relations are valid provided that the effective spectral index is in the range
$-3<n<1$. The effective index on typical galaxy scale for a scale-invariant initial spectrum
is indeed approximately $-2$ (Gunn, 1987, Silk, 1999, Chap.3) and that corresponds to $\gamma'\simeq 2$.

\subsubsection{Adiabatic contraction factor}

In the primeval version of TCV (LS1) we asssumed that the baryonic component when
it reaches virial equilibrium
is  completely done of a collisionless star fluid. That may be considered as an extreme assumption
we have now realistically to soften by a parametrization.
We have to assume that when the main relaxation process ends the
transformation of gas into stars does not be completed. 
 To take into account this physical condition we
refer to the fluid approximation used by Klar \& M\H{u}cket (2008) (hereafter KM8)
in order to follow the dynamical coupling of both components the DM one
and that of baryonic gas. In this approach the gas is allowed to undergo
cooling processes and the consequence of its dissipation on DM is
an adiabatic contraction which, in turn, has a back-reaction onto the
gas dynamics. The assumption is made that the DM is dynamically
reacting fast enough on any change of the overall gravitational potential
via processes of violent relaxation kind.

Two cases are taken into account in KM8: i) the initial DM profile 
follows a NFW (1996) profile, ii) the DM distribution is politropic.
Even if the approach appears very crude for our context
(e.g., the whole baryon mass is in gas with a ratio of baryons over DM equal to
0.25) we may conclude 
from both cases analized  that the effect may be parametrized by
introducing in TCV a mean linear contraction factor
$c= \bar{r}_{B}/\bar{r}_{D} < 1/10$.

\subsection{Interaction term}
As usually done, the Clausius Virial
trace normalized to $G M_B^2F/a_D$ is given by:

\begin{equation}
\label{norvi}
\tilde{V}_B=-\frac{\nu_{\Omega B}}{ x}-\frac{\nu_V(x)}{x}
\end{equation}

Then we have to perform the interaction coefficient $\nu_V $
(Eqs. \ref{CoeIn}, \ref{Coe}) for different  
King's and DM concentrations in the $d$-range $0\div1$.
The results are collected in the Tables of Appendix together with
the explicit formula of $V_B$.

\begin{figure}[ht]
\begin{center}
\includegraphics[width=1.00\textwidth]{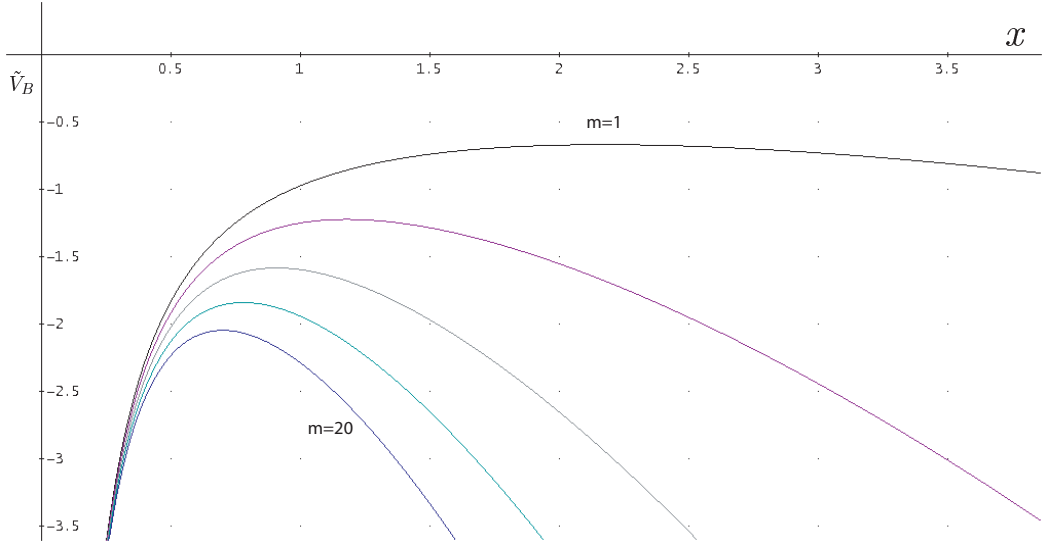}
\end{center}

\caption{Normalized Clausius Virial $\tilde{V}_B$ vs. the ratio of virial radii of the two components:
 $x~(=a_B/a_D)$, for $C_B=C_D=10$
and $m$ in the range $1\div 20$.}

\label{secco1_2}
\end{figure}

In Fig.(\ref{secco1_2}) its typical trend is shown in the case: $C_B=10,~ C_D=10,~ d=0.5,~ m=1\div 20$. 

It is to be remarked that also in this non-linear approximation, at good extent, the ratio:
\begin{equation}
\label{nu}
\nu'_V=\frac{\nu_V(x)}{m x^{3-d}}\simeq const.
\end{equation}
exactly as in the linear formulation 
of TCV. That simplifies enormously the translation of the linear approximation 
into the non-linear one.
As the first consequence is that the Clausius Virial maximum (CVM) appears again at:
\begin{equation}
\label{at}
x_{M}=\Big( \frac{\nu_{\Omega B}}{\nu'_V}
\frac{1}{(2-d)}\frac{M_B}{M_D}\Big)^{\frac{1}{3-d}} 
\end{equation}

Moreover $\widetilde{M_D}$, which is the fraction of $D$ matter 
exerting
dynamical effect on $B$ according to Newton's first theorem, becomes at a good extent,
again as in the linear case:
\begin{equation}
\label{Mtilde}
\widetilde{M_D}\simeq M_D (\frac{a_B}{a_D})^{3-d}
\end{equation}

The same mass fraction normalized to $M_B$ becomes:
\begin{equation}
\label{mtilde}
\widetilde {m}=mx^{3-d}=\widetilde{M_D}/M_B
\end{equation}

Owing to Eq.(\ref{at}), when we consider the special configuration at the
maximum, the normalized mass fraction is:
\begin{equation}
\label{nmft}
\widetilde{m}_M=\frac{\nu_{\Omega B}}{\nu'_V}\frac{1}{2-d}
\end{equation}
which is independent of the mass ratio $m$.

Then total mass inside the B-structure which exerts a dynamical effect on $B$, at CVM becomes:
\begin{equation}
\label{mass}
M_{dyn}= M_B+\widetilde{M_D}=M_B(1+\widetilde{m}_M)
\end{equation}

\subsection{Energy equipartition}
The presence of Clausius' virial maximum means the
virial energy equipartion at $x_M$. It means:
\begin{eqnarray}
\label{equi}
	\Omega_{B}\simeq V_{BD}
\end{eqnarray}

Using the definition of masses (Eq.\ref{Zmass}) by means of their
$(\nu_u)_{M}$ coefficients (\ref{cnum}), energy equipartition translates into the link between
the two central
densities\footnote{Due to the adopted formalism of sect. 5, central density
means scale radius density value.} as follows:

\begin{equation}
\label{lro}
\rho_{oB}\simeq\frac{\nu^{'}_{V}}{\nu_{\Omega B}}\frac{\left(\nu_{D}\right)_{M}}
{\left(\nu_{B}\right)_{M}}\rho_{oD}\frac{1}{\bar{x}^d_M};\\\  \bar{x}_M=cx_M
\end{equation}
where we have taken into account also the adiabatic contraction (see, KM8, subsect.6.1.1) by introduction of
the parameter $c$.

From the physical point of view Eq.(\ref{lro}) means a strict link between the two gravitational
potential wells of baryons and of $DM$ at the special virial configuration corresponding to CVM.

\subsection{Light vs. DM halo}
The King's model relationships allow us to link easily the DM potential well with
the light quantities of the baryonic component in the following way.
The central mass density of the halo, which defines how deep is the corresponding potential well,
is linked to $\rho_{oB}$ (Eq.\ref{lro}). In turn the ratio between the two Eq.(\ref{masD},
\ref{masP}) gives:
\begin{eqnarray}
\label{mPmD}
	\rho_{oB}=\frac{\Sigma\left(0\right)z_{o}}{\left(1-z_{o}\right)^{2}\pi r_{c}}\left[\frac{1}{z_{o}}\cos^{-1}z_{o}-\left(1-z_{o}\right)^{\frac{1}{2}}\right]
\end{eqnarray}

From the other hand Eq.(\ref{masP}) reads: $\Sigma(0)\cdot \frac{k_L}{k_M}=I(0)$.
Then by Eqs.(\ref{mPmD}, \ref{lro}) we obtain how the central surface 
brightness in flux, $I(0)$, links to DM potential well:

\begin{eqnarray}
\label{lumas}
	I(0)= F(z_o)\pi r_{c}(1-z_o)^2
	\frac{\nu^{'}_{V}}{\nu_{\Omega B}}\frac{\left(\nu_{D}\right)_{M}}
	{\left(\nu_{B}\right)_{M}}\rho_{oD}\frac{1}{\bar{x}_M^d}\frac{k_L}{k_M}\\
	\nonumber	
	F(z_o)=\frac{1}{z_o\left[\frac{1}{z_{o}}\cos^{-1}z_{o}-\left(1-z^{2}_{o}\right)^{1/2}\right]}
\end{eqnarray}
According to Eqs.(\ref{limL},\ref{flx}), it is by definition :
\begin{eqnarray}
\label{ref}
	L\left(X_{e}\right)=\frac{1}{2}\pi r^{2}_{c}k_{L}F_{L}\left(X_{t}\right)
\end{eqnarray}
the solutin of which gives $X_e$, the square of effective radius normalized to $r_c$.
On the other hand the following relationship for total luminosity holds:

\begin{eqnarray}
\label{ltot}
	L_{tot}=2\pi r^{2}_{e}I_{e}=\pi r^{2}_{c}k_{L}F_{L}\left(X_{t}\right)
\end{eqnarray}
and then:
\begin{eqnarray}	 
\label{ief}
I_{e}=\frac{1}{2X_e}k_{L}F_{L}\left(X_{t}\right)
\end{eqnarray}

The ratio between Eq.(\ref{ief}) and:
\begin{eqnarray}
\label{io}
	I\left(0\right)=k_{L}\left[1-z_o\right]^{2}
\end{eqnarray}
immediately yields:
\begin{eqnarray}
\label{ie}
I\left(0\right)=2I_{e} X_e\frac{1}{F_{L}\left(X_{t}\right)}\left[1-z_o\right]^{2}
\end{eqnarray}
Inserting it into Eq.(\ref{lumas}) we obtain how the quantity of light given by, $I_e$, depends on the DM potential well: 
\begin{eqnarray}
\label{iec}
 I_{e}=
 F(z_o)\pi r_c
	\frac{\nu^{'}_{V}}{\nu_{\Omega B}}\frac{\left(\nu_{D}\right)_{M}}
	{\left(\nu_{B}\right)_{M}}\rho_{oD}\frac{1}{\bar{x}^d_M}\frac{k_L}{k_M}
\frac{F_{L}\left(X_{t}\right)}{2X_e}
\end{eqnarray}
That is one of the main relationships the TCV yields in order to understand
the physical tilt-mechanism. We will come back later (sect.8). 
\begin{figure}[ht]
	\begin{center}
	\includegraphics[width=1.00\textwidth]{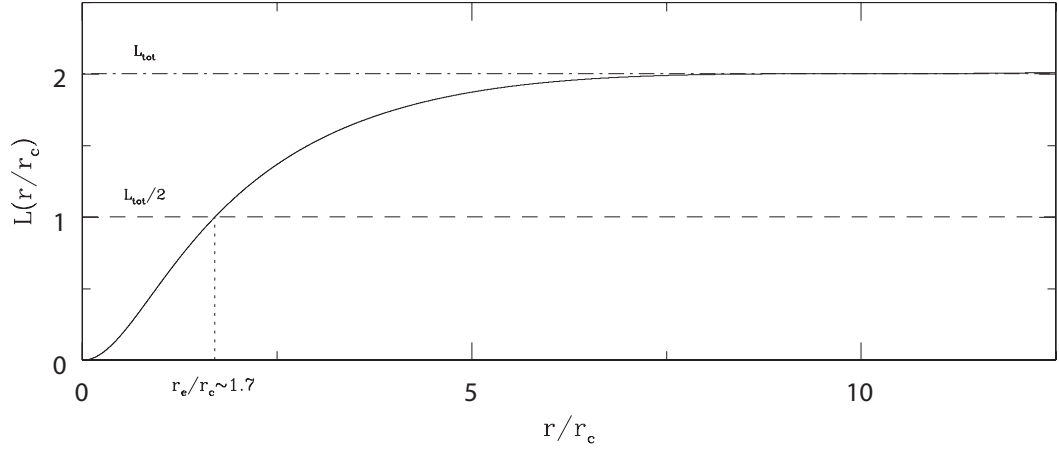}
	\caption{Solution of Eq.(\ref{ref}) in order to find $r_e/r_c$, in the case $r_t/r_c=10$, of $B$ King's component.}
	\label{fig:raggioe}
	\end{center}
\end{figure}

 \section{Theoretical FPs}
 
 To find the theoretical FPs in the present non-linear theory approximation (a $B$ King model embedded 
 into a $D$ power law halo) becomes easy
 due to some reductions of this approach to the linear one. Indeed as soon as the 
 condition (\ref{nu}) holds the whole main linear formalism (LS1) may be recovered.
 Then there are two ways in order to write down the theoretical equation of FP.

 \subsection{Main way}
 Briefly speaking, the physical reason for the existence of the FP lies into the existence of a maximum
 in the Clausius virial energy which is able to divide it in about two equal amounts: the self-potential energy
 of the baryonic component and the tidal potential energy due to the fraction of DM halo which has dynamical effect
 on it. On this special virial configuration the following relation holds:
 
 \begin{equation}
 \label{piano}
  \frac{1}{2}M_B \langle \sigma^2 \rangle = -V_{BD}=\nu'_V GM_BM_DF \frac{a_B^{2-d}}{a_D^{3-d}}
 \end{equation} 
 Extracting $a_M$, i.e., the virial dimension of $B$ component at the maximum which directly 
 links to $r_e$, the FP springs up: 
 \begin{equation}
\label{approx}
a_M\simeq \left( \frac{\frac{1}{2}M_B\frac{\sigma_o^2}{k_v} a_D^{3-d}}
{\nu'_V GM_B M_D F}\right)^{\frac{1}{2-d}};\ \langle\sigma^2\rangle=\frac{\sigma_o^2}{k_v}
\end{equation}
 
 Factorizing the Eq.(\ref{approx}) as:
 \begin{equation}
\label{FPN}
r_e\sim \sigma_o^{\frac{2}{2-d}}a_D^{\frac{3-d}{2-d}}m^{-\frac{1}{2-d}} M_B^{-\frac{1}{2-d}};\ r_e=\frac{k_R}{F} \frac{a_M}{\nu_{\Omega B}}
\end{equation}
it follows immediately:

\begin{eqnarray}
\left\{
\label{guess}
\begin{array}{l}
\sigma_o^A\equiv \sigma_o^{\frac{2}{2-d}}\\
\label{guesI}
I_e^B\sim a_D^{\frac{3-d}{2-d}}m^{-\frac{1}{2-d}} M_B^{-\frac{1}{2-d}}
\end{array}
\right.
\end{eqnarray}
$k_R,k_v$ are the usual coefficients for kinematic and density galaxy distributions.
On the contrary of the linear approximation, the present model allows us not only to 
give explicitely the numerical factor in the second relation of (\ref{guess}) but
to understand deeply the physical meaning of the previous factorization.

\subsection{The most easy way}
The most easy way to obtain from TCV the theoretical FPs is the following.
From two-component virial equation (Eq. \ref{vireq}):
\begin{equation}
T_B=-\frac{1}{2}\Omega_B-\frac{1}{2}V_{BD}
\end{equation}
by remembering Eq. (\ref{vir}), the definitions of $T_B, \sigma_o ^2$ and $r_e$ in Eqs. (\ref{piano}, 
\ref{approx}, \ref{FPN}), it follows:
\begin{equation}
\label{otw}
\frac{\sigma_o^2}{k_v}=\frac{GM_B}{r_e}k_R + \frac{\nu_V}{\nu_{\Omega B}} \frac{G\widetilde{M}_D}{r_e}k_R 
\end{equation}

and then Eqs. (\ref{Mtilde},\ref{nmft}), we obtain:
\begin{equation}
\label{massa}
M_B=\sigma_o^2 r_e c_2 \left[\frac{2-d}{3-d}\right]
\end{equation}
where $c_2=(Gk_Rk_v)^{-1}$
turns out to be a constant, if we assume that homology
holds for kinematic and density distributions of elliptical galaxies. 
Here we also assume that $c_2$ comes out from King's model as given by
Bender et al. (1992, Fig.5) in the case of isotropic velocity dispersion and with
an unchanged distribution even if the $B$ King's component is now embedded in a DM halo.

When Eq.(\ref{massa}) is divided by $L=c_1 I_e r_e^2~~$ (here $c_1=2\pi)$, due to
the special configuration of CVM, then $L/M_B\sim M_B^{-\frac{1-d}{3-d}}$. So
the theoretical FP arises in the form:
\begin{equation}
\label{FPi}
r_e=(c_2c_3)^{\frac{A}{2}} c_1^{B} (L^o)^{-B}(M_B^o)^{-\frac{A}{2}} \sigma_o^{A} I_e^{B}
\end{equation}
where:
\begin{eqnarray}
A=\frac{2}
{2-d};\
B=-\frac{3-d}{2(2-d)}\\
c_1=2\pi;~~c_2=c_2(C_B);\
c_3=\left(\frac{2-d}{3-d}\right);
\end{eqnarray}
$L^o$ and $M_B^o$ are luminosity and mass of one elliptical galaxy choosen in order to calibrate the plane. 

\subsection{To test the theoretical FPs}

\begin{figure}[!ht]
	\begin{center}
\includegraphics[width=1.00\textwidth]{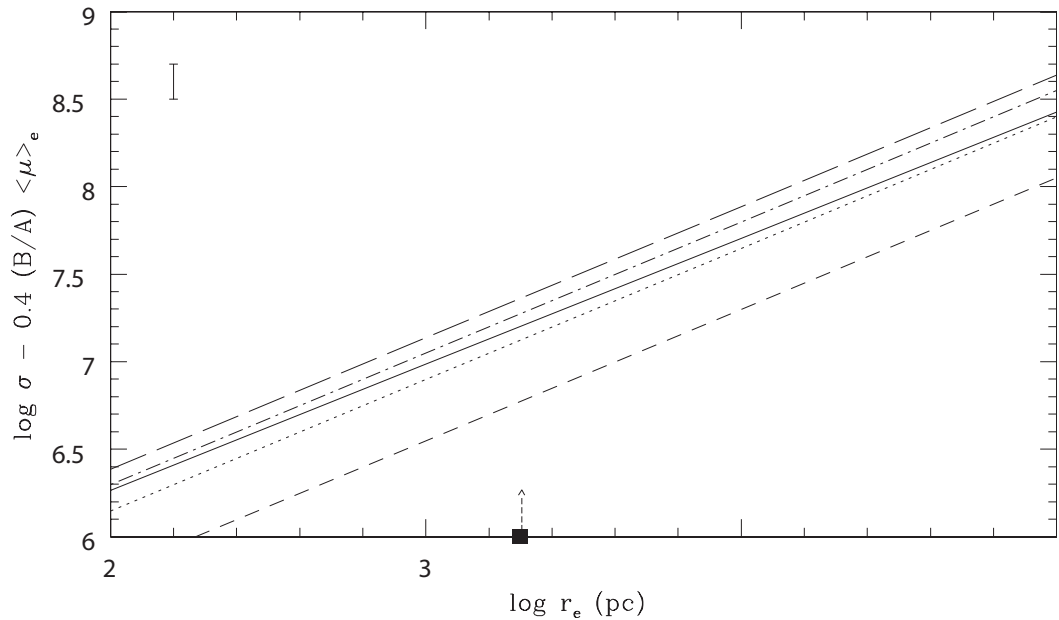}
	\caption{Theoretical FPs as outputs of Eq.(\ref{FPi}) for $d=0.5$ 
	with different calibration $M_B^o/L^o=15,10,5,1$ (from top to down, long-dashed,
dot-dashed, dotted and short-dashed lines, respectively) in $r_G$ band. The fit-equation of early-type galaxies 
obtained by Djorgovski \& Davies (1987), in the Lick $r_G$ band, is also plotted as comparison (solid line) with the median vertical error bar
in the left corner. A vertical arrow signs the reference galaxy used for the theoretical calibration.}
	\label{fig:FP1}
	\end{center}
\end{figure}

\begin{figure}[ht]
	\begin{center}
\includegraphics[width=1.00\textwidth]{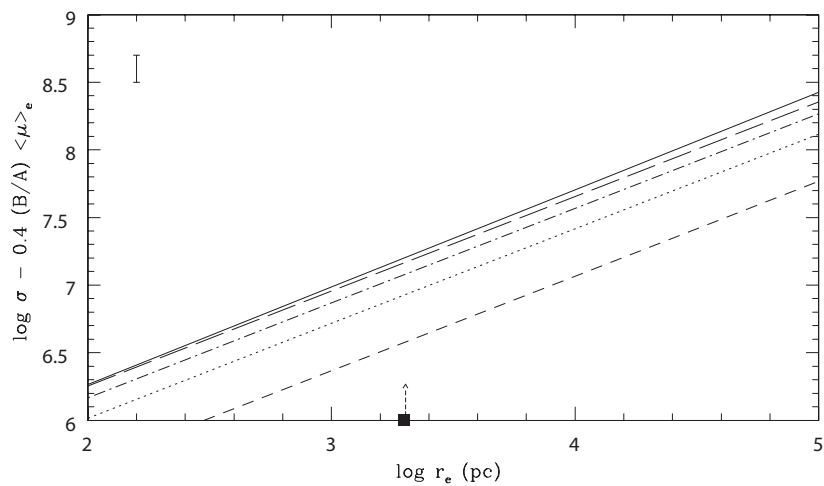}
	\caption{As in Fig.(\ref{fig:FP1}) with $d=0.6$ calibrated with $M_B^o/L^o=15,10,5,1$ (from top to down, with the same previous line types) and the fit of Djorgovski \& Davies (1987) (solid line), in Lick $r_G$ band, as comparison with the median vertical error bar
in the left corner. A vertical arrow signs the reference galaxy used for the theoretical calibration.}
	\label{fig:FP2}
	\end{center}
\end{figure}

We compare the theoretical FPs produced by Eq.(\ref{FPi}) 
with that obtained by Djorgovski \& Davies (1987) (hereafter, $DD87$; see also, Kormendy \& Djorgovski, 1989) 
by fitting the observations (in the Lick $r_G$ band):
\begin{equation}
\label{Djor}
\log r_e= 1.39 (\log \sigma_o+0.26 \langle\mu\rangle_e)-6.71
\end{equation}
We plot in Figs.(\ref{fig:FP1}), (\ref{fig:FP2})
the edge-on FPs as follows:  
\begin{equation}
\label{FPLot}
\frac{\log r_e+ const}{A}=\log \sigma_o-0.4\frac{B}{A}\langle\mu\rangle_e
\end{equation}
where the values of the parameters are given in Tab.(1).
\begin{table}
\centering
\begin{tabular}{|c|c|c|c|c|}
\hline
\hline
$d$ &$M^{o}_{B}/L^{o}$ & $A$ & $B$ & $const$ \\
\hline
\textbf{0.5}	&	1	&	1.33	&	-0.83	&	5.73	\\
"	&	5	&	1.33	&	-0.83	&	6.20	\\
"	&	10	&	1.33	&	-0.83	&	6.40	\\
"	&	15	&	1.33	&	-0.83	&	6.52	\\
\hline
\textbf{0.6}	&	1	&	1.43	&	-0.86	&	6.10	\\
"	&	5	&	1.43	&	-0.86	&	6.60	\\
"	&	10	&	1.43	&	-0.86	&	6.81	\\
"	&	15	&	1.43	&	-0.86	&	6.94	\\
\hline
\hline
\end{tabular}
\caption{Values of parameters which enter into Eq.(\ref{FPLot}).}
\label{tab:const}
\end{table}
For both figures the theoretical FPs from Eq.(\ref{FPi}) are plotted: i)  for $d=0.5$ (Fig.\ref{fig:FP1}) and $M_B^o/L^o=15,10,5,1$ (from top to down, long-dashed,
dot-dashed, dotted and short-dashed lines, respectively)
 and: ii)
for $d=0.6$ ( Fig.\ref{fig:FP2}) and ratios $M_B^o/L^o=15,10,5,1$ (from top to down of, with the same previous line types) and  the $DD87$
(solid line) is also shown as comparison. 

To calibrate the theoretical plane (\ref{FPi}) we use $L^o$ and $M_B^o$ of the 
elliptical galaxy $PGC045032$ $(1300.4+2807)$ of Coma Cluster
which has been fitted by King's profile (Oemler, 1976) as follows:

\begin{eqnarray}
M_o'=-16.89;\ \log(r_t)=1.10 (Kpc);\ \log(r_c)=0.10 (Kpc)\
\end{eqnarray}

By conversion of $M_o'=-2.5 \log(I(0)r_c^2) + const$ (in $V$ band) into: $I(0)=3.08\cdot 10^8 L_{\odot}/(Kpc)^2$
and then using Eqs.(\ref{io},\ref{ie},\ref{ltot}) of King's model ($r_t/r_c=10;X_e= (1.70)^2;~F_L(X_t)=2$), we obtain
$k_L=3.79\cdot 10^8 L_{\odot}/(Kpc)^2
; I_e=1.31\cdot L_{\odot}/(Kpc)^2$ 
and the total luminosity : $L^o=3.78\cdot 10^9 L_{\odot}$.

To transform the photometric data of our reference galaxy from the Johnson\footnote{Actually we used 
the UBVRI bands for the combined Johnson-Cousins-Glass system and the solar corresponding
values as given by Binney \& Merrifield (1998, Chap.2, pg.53).} BVR system into Lick $r_G$ band pass
we use the Djogovski (1985) transformations: $(V_J-r_G)=0.031+0.681(B-V)_J$, assuming a mean $(B-V)_J\simeq 1$, 
$(r_G)_{\odot}=4.43$ and 
$\log I_e=-0.4( \langle\mu\rangle_e)-26)$. At this level of first approximation comparison we do not
consider how could change $r_e$ and $\sigma_o$ of our reference galaxy by changing the photometric
color band.   

The comparison looks fairly well. Inside the vertical median bar the edge-on observed FP of $DD87$ lies  between the two
theoretical FPs corresponding to $M^{o}_{B}/L^o=10$ and  
$M^{o}_{B}/L^o=5$, in the case $d=0.5$ (Fig.\ref{fig:FP1}). The  theoretical values $A^*,B^*, -0.4B^*/A^*$ 
are $1.33,-0.83,0.25$ and the observed ones are: $A=1.39\pm 0.15, B=-0.90\pm 0.10,-0.4B/A= 0.26\pm0.06$. Then inside the 
error bars the two results coincide even if the theoretical straight lines 
turn out to be a little bit steeper in respect to the observed one. The other case $d=0.6$ (Fig.\ref{fig:FP2}) looks better
from the slopes point of view: the
theoretical values become: 
$A^*=1.43, B^*=-0.86, -0.4B^*/A^*=0.24$ in very good agreement with the fit result. 
The two slopes,  theoretical and observed, are then nearer, even if 
an higher value of $M^{o}_{B}/L^o=15$ seems to be preferred without exclusion of the case $M^{o}_{B}/L^o=10$.
Concluding, from this first approximation comparison both cases $d=0.5, d=0.6, M^{o}_{B}/L^o=10$ are acceptable 
inside the error bars of observed fit with an $M^{o}_{B}/L^o=5$ to be preferred in the case $d=0.5$ from
the reference galaxy (vertical arrow) forwards. On the contrary
the case $d=0.6$ seems to request an higher ratio $M^{o}_{B}/L^o$ at least of $10$.

\section{The most relevant way}
The most relevant way to understand the physical meaning of FP occurs as soon as 
we wish to translate the theoretical FP described by the relationship (\ref{FPN}) into the $\kappa$-space (Bender et al.
1992). Now we know the second relation (\ref{guesI}) as an equation by King's model, that is
Eq.(\ref{iec}). The (\ref{guesI}) tells us how $I_e$ has to scale  with $M_D$ and $M_B$. We know already
it from LS1:
\begin{eqnarray}
\label{scalie}
	I_{e}\sim m^{i}M^{I}_{B}\sim M^{i}_{D};\ i=I=2\frac{\gamma'-\left(3-d\right)}{\gamma'\left(3-d\right)}
\end{eqnarray}

but it is Eq.(\ref{iec}) which allows us to understand deeply why this scaling law has to be
followed.
We can indeed to recover the (\ref{scalie}) by Eq.(\ref{iec}) as soon as
we remember that:
\begin{equation}
\label{cosm}
\rho_{oD}\sim M_D^{-(\frac{3}{\gamma'}-1)}
\end{equation}
according to subsct.6.1.
By using Eq.(\ref{at}) it follows:
\begin{equation}
r_c=\frac{1}{C_B}x_M a_D
\end{equation}
where from cosmology (Eq. \ref{adM}), $a_{D}\sim M^{\frac{1}{\gamma'}}_{D}$.

Remembering that:
\begin{eqnarray}
k_L/k_M=\frac{L}{M_{B}}=M^{-\alpha}_{B}=M^{-\frac{1-d}{3-d}}_{B}	
\end{eqnarray}
we obtain, at fixed $C_B$:
\begin{eqnarray}
I_{e}\sim r_{c}\rho_{oD}\left(\frac{a_{D}}{a_{B}}\right)^{d}\frac{L}{M_{B}}
\sim M^{\frac{1}{3-d}}_{B}\cdot M^{-\frac{1}{3-d}}_{D}\cdot M^{\frac{1}{\gamma'}}_{D}\\
\nonumber
\cdot M^{-\left(\frac{3}{\gamma'}-1\right)}_{D}\cdot M^{\frac{d}{\gamma'}}_{D}
\cdot M^{-\frac{d}{3-d}}_{B}\cdot M^{\frac{d}{3-d}}_{D}\cdot M^{-\frac{d}{\gamma'}}_{D}
\cdot M^{-\frac{1-d}{3-d}}_{B}\\
\nonumber
\sim M^{x}_{D}\cdot M^{y}_{B}
\end{eqnarray}

with the exponents
\begin{eqnarray}
x=-\left(\frac{3}{\gamma'}-1\right)+\frac{d}{\gamma'}+\frac{d}{3-d}-\frac{d}
{\gamma'}-\frac{1}{3-d}+\frac{1}{\gamma'}\\
\nonumber
=2\frac{\gamma'-\left(3-d\right)}{\gamma'\left(3-d\right)}=I\\
y=\frac{1}{3-d}-\frac{d}{3-d}-\frac{1-d}{3-d}=0
\end{eqnarray}
which matchs exactly (\ref{scalie}).

The physical meaning is the following: light does not follow the visible matter
because it depends, via stars, on the deepness of the gravitational potential well
which is determined by the two central densities of DM and baryons linked together
by the equipartion of virial energy, that is by the Eq.(\ref{lro}).

\subsection{$I_e$- scaling}

If we take into account two galaxies corresponding to two different baryonic masses
$(M_{B}^o,M_B)$ but characterized by the same $m=M_D/M_B=M^o_D/M^o_B$, $C_B$ and $C_D$
(i.e., the same CVM, $x_M=x_M(d,m,C_B,C_D))$, $I_e$ has to scale as:

\begin{equation}
\label{scie}
I_e/I_{e}^o= (M_B/M_{B}^o)^I
\end{equation}
which reduces to:
\begin{equation}
\label{sie}
I_e/I_{e}^o= (M_B/M_{B}^o)^{-\alpha}
\end{equation}
if we are in the typical mass range of $M_D\simeq 10^{11}M_{\odot}$, that is $\gamma'\simeq 2$.

\subsection{$r_e~,\sigma_o$- scaling}
From Eq.(\ref{massa}) we know that:
\begin{equation}
\sigma_o^2=\frac{M_B}{r_ec_2c_3}
\end{equation}

where $r_e$ scales in turn with $M_B$ as follows:
\begin{equation}
r_e\sim M_B^{1/2}k_M^{-1/2}
\end{equation}
due to the King's model relationship, $k_Mr_c^2\sim M_B$, at fixed $C_B$.
But at the Clausius' virial maximum configuration, $r_e$ has to scale as (LS1):
\begin{equation}
r_e\sim m^rM_B^R;~~~~~r=I/2;~~R=1/\gamma'
\end{equation}
For the two galaxies considered before, at fixed $x_M$, it follows that:
\begin{equation}
\label{sre}
r_e/r_e^o=\left(M_B/M_B^o\right)^{1/\gamma'}
\end{equation}
as soon as $k_M^{1/2}\sim M_B^{1/2-1/\gamma'}$, and:
\begin{equation}
\label{ssigm}
\left(\sigma_o/\sigma_o^o\right)^2=(M_B/M_B^o)^{\frac{\gamma'-1}{\gamma'}}
\end{equation}

\section{Theoretical {\it tilt} equation in $\kappa$-space}
Following Bender et al.(1992) we have to build up the theoretical {\it tilt}
equation in the $\kappa$-space.
From Eqs.(\ref{scie},\ref{sre},\ref{ssigm}) we obtain for $\kappa_1$ and $\kappa_3$, respectively:
\begin{eqnarray}
\kappa_1=(\log\sigma_o^2+\log r_e)/\sqrt{2}\\
\nonumber
   =(\log(\sigma_o^o)^2+\log r_e^o)+\log(M_B/M_B^o))/\sqrt{2}\\
   \label{klink}
 \log(M_B/M_B^o)= \sqrt{2}\kappa_1- [\log(\sigma_o^o)^2+\log r_e^o]
\end{eqnarray}
and
\begin{eqnarray}
\kappa_3=(\log\sigma_o^2 -\log I_e-\log r_e)/\sqrt{3}\\
\nonumber
   =(\log(\sigma_o^o)^2-\log I_e^o-\log r_e^o)+\alpha_e\cdot \log(M_B/M_B^o))/\sqrt{3}\\
 \alpha_e=\alpha_e(n_{rec}, d)=(\frac{\gamma'-2}{\gamma'}-I(\gamma', d))
 \end{eqnarray}
Inserting the link (Eq.( \ref{klink})) between $(\kappa_1, \kappa_3)$, we obtain the {\it tilt}
equation:
\begin{equation}
\label{tkspae}
\kappa_3=\alpha_e\cdot \sqrt{2/3}~~ \kappa_1+ \frac{(1-\alpha_e)\log(\sigma_o^o)^2-(1+\alpha_e)\log r_e^o-
\log I_e^o}{\sqrt{3}}
\end{equation}
The last equation tells us how the FP becomes degenerate
in respect to the cosmology. Indeed, as soon as we are in the typical galaxy  mass range,
so that, $\gamma'\simeq 2$ (Gunn, 1987, Silk, 1999, Chap.3), then $-I\rightarrow \alpha$, according to Eq.(\ref{sie}), and $\alpha_e\rightarrow \alpha$
which is depending only on $d$, i.e., on the halo mass distribution.
In this range the galaxy FP becomes:
\begin{equation}
\label{tkspa}
\kappa_3=\alpha\cdot \sqrt{2/3}~~ \kappa_1+ \frac{(1-\alpha)\log(\sigma_o^o)^2-(1+\alpha)\log r_e^o-
\log I_e^o}{\sqrt{3}};~~\alpha=\frac{1-d}{3-d}
\end{equation}

The last equation shows that: if $d=1\rightarrow\alpha=0$ and then $\kappa_3=const.$,
which means the {\it tilt} disappears!

\subsection{Calibration}
\begin{figure}[ht]
\begin{center}
\includegraphics[width=1.00\textwidth]{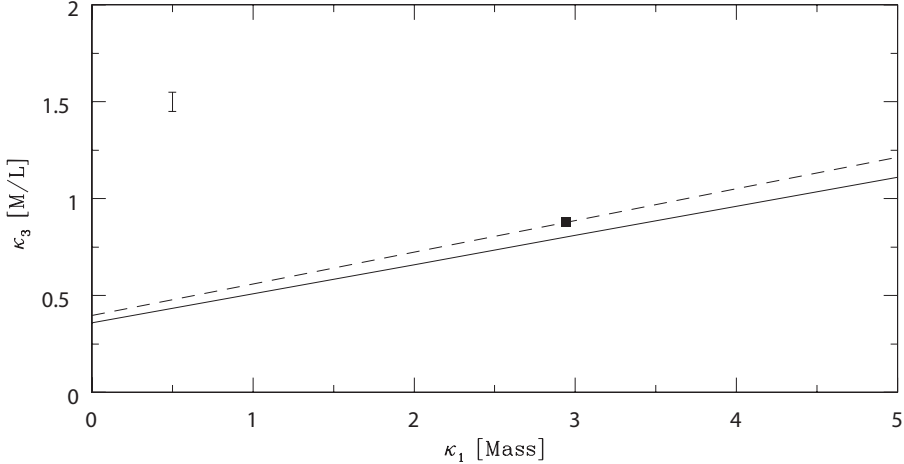}
\end{center}

\caption{Comparison between the theoretical tilt-equation (Eq.(\ref{eqkt}); dashed line) (d=0.5) and that
derived by the observational fit in B-band of Burstein et al.(1997) (solid line). 
The discrepancy between the two straight lines is $\Delta \kappa_3= 0.07$ at the reference galaxy 
used for calibration (signed by a filled square) less than 2 times the assigned tightness of FP, $\sigma(\kappa_3)=0.05$,
shown
on the left corner.} 
\label{k1k3_05}
\end{figure}

Again as in subsect. 7.3, we calibrate the FP by the elliptical galaxy $PGC045032$ $(1300.4+2807)$ of
Coma cluster ($z_C=0.023$) 
which has been fitted by King's profile (Oemler, 1976), characterized by:
\begin{equation}
\label{set}
C_B=10;~F(z_o)=0.7294;~X_e= (1.70)^2;~F_L(X_t)=2
\end{equation}
We choose for all the galaxies of the theoretical plane the
same Clausius' virial maximum configuration, corresponding to:
\begin{eqnarray}
\label{set}
C_D=C_B=10;
~m=10;~~~d=0.5;~~~ \nu'_V=0.0697;~~~
\nu_{\Omega B}=0.9039\\
\nonumber
(\nu_D)_M=0.5463;~~~(\nu_B)_M=0.0103;\
x_M= 0.9;
\end{eqnarray}
If our reference galaxy has a ratio: $k_L/k_M=1/5$, $B-V\simeq 1$,
we obtain, in $B$ band:
\begin{eqnarray}
\log(\sigma_o^o)^2=3.83\\
\nonumber
\log I_e^o=1.98 (\rightarrow I_e^o=95.2 L_{\odot}/pc^2;\ L^o=2.61\cdot10^9 L_{\odot})\\
\nonumber
\log r_e^o=0.33 (\rightarrow r_e^o=2.14 Kpc)\\
\kappa_3^o=0.88, ~~~\kappa_1^o=2.94
\end{eqnarray}
Then the theoretical {\it tilt} equation (\ref{tkspa}) in $\kappa$- space ( $d=0.5\rightarrow \alpha =0.2; \gamma'\simeq 2$) becomes:
\begin{equation}
\label{eqkt}
\kappa_3=0.16 \kappa_1+0.40
\end{equation}
to be compared with that given in $B$ band by Burstein et al.(1997):
\begin{equation}
\label{eqkB}
\kappa_3^*=0.15 \kappa_1^*+0.36
\end{equation}
The two equations are plotted in Fig.\ref{k1k3_05}.
The difference in $\kappa_3$ at fixed $\kappa_1$ (Fig.\ref{k1k3_05}) turns out to be: $\Delta \kappa_3= 0.07$, a little bit
over the assigned FP tightness: $\sigma(\kappa_3)=0.05$.

\begin{figure}[ht]
\begin{center}
\includegraphics[width=1.00\textwidth]{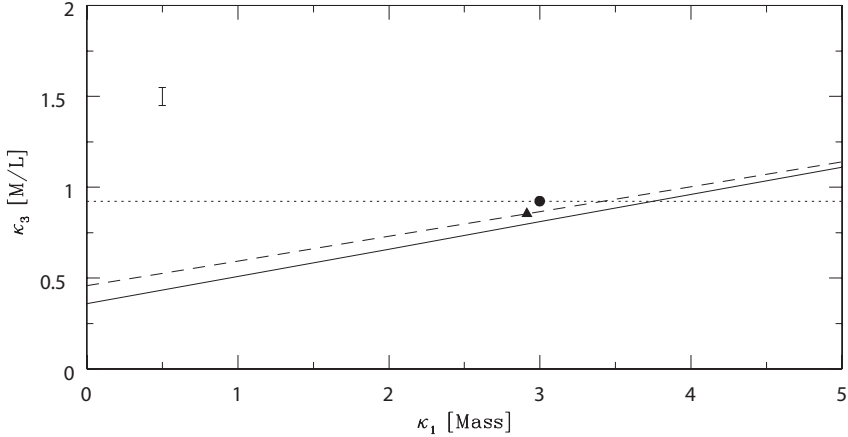}
\end{center}

\caption{Comparison between the theoretical tilt-equation (Eq.(\ref{tkspa}); dashed line) ($d=0.6$; $\kappa_3=0.14\kappa_1+ 0.46$) and that
derived by the observational fit in B-band of Burstein et al.(1997) (solid line). 
The discrepancy between the two straight lines is $\Delta \kappa_3= 0.06$ at the reference galaxy 
used for calibration (signed by a filled triangle, $\kappa_1^o=2.91; \kappa_3^o=0.85$), about the assigned tightness of FP, $\sigma(\kappa_3)=0.05$ shown
on the left corner. The limit case $d=1$ for which the tilt desappears, is also plotted (dotted line) with the relative
position of the calibration galaxy (filled circle).} 

\label{k1k3_106}

\end{figure}
It should be noted that the DM halo of our reference galaxy turns out to be 
characterized by:
\begin{equation}
M_D=m M_B^o= 1.3\cdot 10^{11} M_{\odot};\ \gamma'\simeq 2;
\end{equation}
Its formation redshift has been calculated better than by Eq. (\ref{redsh}), using the subroutine 
of Navarro et al. (1997) in which
$z_F$ is precisely defined as the $z$ at which half of final mass is in progenitors
more massive than $1\%$ of the final mass. At this $z_F=3.2$ ($h=0.75$) the corresponding overdensity (in
units of the critical density at $z=0$) becomes
$\delta_c=2.42\cdot 10^5$ which corresponds to the value of 
central density: $(\rho_{oD})_{NFW}\simeq 3.3\cdot 10^{-2}M_{\odot}/pc^3$,
for NFW-profile
used by Navarro et al. (1997).
\footnote{For this profile, central density
means density at about one half of scale radius.}
Rescaling this value to the cored power-law profile of Eq.(\ref{dmp}) ($d=0.5$) by a factor of about $1.17$, 
to match the theoretical value of $I_e$
given by Eq.(\ref{iec}) we need of 
contraction factor $c\simeq 1/15$. That appears consistent with
the limits found by 
Klar \& Muecket (2008) (sect. 6), considering that in this context most of
baryonic matter is in stars.

\subsection{Discussion and conclusion}

Both the comparisons of theoretical FPs either with that of Diorgovski \& Davies (1987) in Lick $r_G$
or that of Burstein et al.(1997) in $B$ band result very satisfactory. Moreover 
the  edge-on theoretical representation, $(\kappa_3, \kappa_1)$, of FP (Fig.\ref{k1k3_106}) appears
to rotate around the calibration point as soon as $d$ increases according to Eq.(\ref{tkspa}). It becomes horizontal when
$d$ becomes equal $1$. In this limiting case: $d=1\rightarrow \alpha=0; \kappa_3=\kappa_3^o$
whichever is $\kappa_1$ and then the FP looses the {\it tilt} (Fig.\ref{k1k3_106}) as already underlined in the previous papers
LS1, LS5. 

The new order of approximation of TCV by the inclusion of
a King-like $B$ component allows us to understand more deeply the physical reason of
the FP {\it tilt}.
The CV theory is born to try to explain the FP (firstly of ETGs)  without breaking the {\it homology+virial equilibrium}. 
When the galaxy system is looked as a whole system a dynamical explanation of the observed {\it tilt} turns out
to be impossible without breaking homology: the exponents $A,B$ become rigorously $2,-1$ due to the virial equilibrium.
The only way to change the exponents is to split the system into two subsystems: one of baryons, the other one of DM
as done by the tensor virial theorem.
Then the double system may gain a new symmetry due to the equipartition of Clausius' virial energy between dark and
visible matter. When that occours, the virial configuration becomes special because its CV energy reaches
a maximum value, i.e., a minimum of its random kinetic energy to obtain equilibrium. This allows to the
exponents of $\sigma_o$ and $I_e$ to  decrease their absolute value if the bulk of baryonic matter lies inside
a dark matter halo distribution like a power law $\rho_D\sim r^{-d}; d\simeq0.5$. But that is not enough.
The reason of the {\it tilt}, at this higher order of approximation of TCV,  appears  more clearly. 
It lies in that: light does not follow the visible matter because the star
formation is due to the two potential wells, of baryons and of DM. Their depths are linked together
by the equipartition of virial energy, that is the relationship given by Eq.(\ref{lro}). 
Then the main conclusion is the following: in our approach the galaxy {\it tilt} is neither due to 
different DM fraction which enters into the dynamical mass of Eq.(\ref{mass}), in order
to increase the observed ratio $M_{dyn}/L$ (i.e., $\kappa_3$) at increasing $M_{dyn}$ (i.e., $\kappa_1$) starting 
from a
fixed mass-luminosity ratio $\Upsilon_*
=M_B/L$ for all galaxies.
That has been already understood by Ciotti et al. , since 1996, who realize a fine-tuning
was invoked. 
Nor it may be
explained by trying to tune $M_{dyn}/L$ by different DM amount, again with a constant $\Upsilon_*$, 
assuming the galaxies are located on the
CV maximum configuration as Valentinuzzi (2006) tried  whithout success (even if
his approach was
substantially different in respect to the TCV). In our approach the fraction of $M_{dyn}$ in DM  
is always constant, as soon as we locate the ETG at a fixed amximum: $x_M=x_M (d, m, C_D, C_B)$. That is:
\begin{equation}
\label{frat}
\left (\frac{\widetilde{M_D}}{M_{dyn}}\right )_M=\frac{1}{1+\frac{\nu'_V}{\nu_{\Omega B}}
(2-d)}
\end{equation}
In this way the: $$\frac {L}{M_B}=\frac{L}{M_{dyn}}(1+ \tilde{m}_M)$$ 
That means the DM amount which enters into $M_{dyn}$ increases  at increasing $M_B$ but
in a way proportional to $M_B$, so that  
any tilt is produced if we start with $\Upsilon_*= const.$. The TCV offers 
the dynamical mechanism in order to change  $\Upsilon_*$ exactly $\sim M_B^{\alpha}$.

The other important answer of TCV is why the FP as a whole appears to be degenerate
in respect to cosmological density perturbation spectrum. Looking at Eq.(\ref{tkspae})
the {\it tilt} looks apparently not independent of cosmology because $\alpha_e$ is depending explicitely by $\gamma'$
which appears also inside the exponent $I$. But in CDM scenario the 
effective index on typical galaxy scale, for a scale-invariant initial spectrum,
is indeed approximately $-2$ (Silk, 1999, Chap.3) and that corresponds to $\gamma'\simeq 2$.
So the $\alpha_e$ degenerates into $\alpha$ which is independent of cosmology. It depends only
on the DM distribution $d$. That is also the reason why the ratio $L/M_B$ is totally independent
of the cosmic perturbation spectrum and of mass ratio $m$ as we have already proved in LS1.
The degeneracy is broken as soon as we look at the projections into the coordinate planes,
as shown in LS1.

Many problems are still open. 
We have considered a unique maximum for all ETG and a unique $C_B$ for the King-like $B$ component. 
What occurs by changing them. What about the other parameters involved?
What happens as soon as we change $\gamma'$?
Moreover we may wonder how the results do change 
moving to a $\Lambda CDM$ scenario, even if 
we expect no significant variations essentially
because the mass variance does not change too much in this last cosmology.
But the main point is: why has the DM distribution which contains the bulk
of baryons follow a density power law of the kind $\rho_D\sim r^{-d}; d=0.5\div 0.6$ in order 
to produce the observed {\it tilt}. Many efforts have been devoted to this problem.
 Theoretical arguments based on dynamics (M\H{u}cket \& Hoeft, 2003) and thermodynamics
(Secco et al., 2007) strengthened by observations lead on this direction
even if numerical simulations seem to prefer $d\simeq 1$ (see, e.g., Bindoni, 2008).
At the moment a definitive answer to this crucial point does not exist.

\section*{Acknowledgements}   We like to thank Roberto Caimmi for
fruitful discussions  and mathematical support, Volker M\H{u}ller
and Jan Peter M\H{u}cket of AIP for their warm hospitality,
constructive comments and very helpful suggestions.
  
\appendix  
\section{Appendix} 
\scriptsize
%\footnotesize

\subsection{King's dimensionless density profile}
\label{sec:KingSDimensionlessDensityProfile}

The King (1962) spatial density profile of Eq.\ref{masD}, in the explicit form, is:
\begin{eqnarray}
\begin{split}
	\rho\left(r\right)=&\frac{k_M}{\pi r_{c}\left[1+\left(r_{t}/r_{c}\right)^{2}\right]^{\frac{3}{2}}}\left[\frac{1+\left(r_{t}/r_{c}\right)^{2}}{1+\left(r/r_{c}\right)^{2}}\right]\cdot\\
&\cdot	\left\{\left[\frac{1+\left(r_{t}/r_{c}\right)^{2}}{1+\left(r/r_{c}\right)^{2}}\right]^{\frac{1}{2}}\cdot\cos^{-1}\left[\left(\frac{1+\left(r/r_{c}\right)^{2}}{1+\left(r_{t}/r_{c}\right)^{2}}\right)^{\frac{1}{2}}\right]-\left[\frac{\left(r_{t}/r_{c}\right)^{2}-\left(r/r_{c}\right)^{2}}{1+\left(r_{t}/r_{c}\right)^{2}}\right]^{\frac{1}{2}}\right\} 
\end{split}
\end{eqnarray}
If we define $\xi_{B}=r/a_{B}=r/r_{t}$ and $C_{B}=r_{t}/r_{c}$, which 
means $C_{B}\xi_{B}=r/r_{c}$, we can write: 
\begin{eqnarray}
	\rho\left(\xi_{B}\right)=\frac{k_{M}}{\pi r_{c}\left[1+C^{2}_{B}\right]^{\frac{3}{2}}}
\cdot\left[\frac{1+C^{2}_{B}}{1+(C_{B}\xi_{B})^{2}}\right]\\
\nonumber	
\cdot\left\{\left[\frac{1+C^{2}_{B}}{1+(C_{B}\xi_{B})^{2}}\right]^{\frac{1}{2}}\cdot\cos^{-1}\left[\left(\frac{1+(C_{B}\xi_{B})^{2}}{1+C^{2}_{B}}\right)^{\frac{1}{2}}\right]-\left[\frac{C^{2}_{B}-(C_{B}\xi_{B})^{2}}{1+C^{2}_{B}}\right]^{\frac{1}{2}}\right\} 
\end{eqnarray}

and:
\begin{eqnarray}
	\rho_{o}=\rho\left(\xi=1/C_{B}\right)=\frac{k_{M}}{\pi r_{c}}\frac{1}{\left[1+C^{2}_{B}
	\right]^{\frac{3}{2}}}\left[\frac{1+C^{2}_{B}}{2}\right]\\
\nonumber
\cdot\left\{\left[\frac{1+C^{2}_{B}}{2}\right]^{\frac{1}{2}}\cdot\cos^{-1}\left[\left(\frac{2}{1+C^{2}_{B}}\right)^{\frac{1}{2}}\right]-\left[\frac{C^{2}_{B}-1}{1+C^{2}_{B}}\right]^{\frac{1}{2}}\right\} 
\end{eqnarray}
Then, the King's profile, normalized to the scale radius density value, is:
\begin{eqnarray}
f_{B}(\xi_{B})=\frac{\rho\left(\xi_{B}\right)}{\rho_{o}}=\frac{2}{1+(C_{B}\xi_{B})^{2}}\\
\nonumber
\cdot\left\{\left[\frac{1+C^{2}_{B}}{1+(C_{B}\xi_{B})^{2}}\right]^{\frac{1}{2}}\cdot\cos^{-1}\left[\left(\frac{1+(C_{B}\xi_{B})^{2}}{1+C^{2}_{B}}\right)^{\frac{1}{2}}\right]-\left[\frac{C^{2}_{B}-(C_{B}\xi_{B})^{2}}{1+C^{2}_{B}}\right]^{\frac{1}{2}}\right\}\cdot\frac{1}{H}
\end{eqnarray}
where:
\begin{eqnarray}
H=\left\{\left[\frac{1+C^{2}_{B}}{2}\right]^{\frac{1}{2}}\cdot\cos^{-1}\left[\left(\frac{2}{1+C^{2}_{B}}\right)^{\frac{1}{2}}\right]-\left[\frac{C^{2}_{B}-1}{1+C^{2}_{B}}\right]^{\frac{1}{2}}\right\}
\end{eqnarray}
%\end{document}
\subsection{DM dimensionless density profile}
\label{sec:DMDimensionlessDensityProfile}
The cored power law that describes the DM density profile is:
\begin{eqnarray}
\rho_{D}(r)=\frac{2\rho_{oD}}{1+(r/r_{oD})^{d}}
\end{eqnarray}
where $r_{oD}$ is the scale radius and $\rho_{oD}$ the density value at the scale radius.
In the usual way, once defined $\xi_{D}=r/a_{D}$ and $C_{D}=a_{D}/r_{oD}$, which means $C_{D}\xi_{D}=r/r_{oD}$, we can write:
\begin{eqnarray}
\rho_{D}(\xi_{D})=\frac{2\rho_{oD}}{1+(C_{D}\xi_{D})^{d}}
\end{eqnarray}
Then the normalized DM profile becomes:
\begin{eqnarray}
f_{D}(\xi_{D})=\frac{\rho_{D}(\xi_{D})}{\rho_{oD}}=\frac{2}{1+(C_{D}\xi_{D})^{d}}
\end{eqnarray}

\subsection{Calculation of Clausius Virial}
\label{sec:CalculationOfClausiusVirial}
We define all the coefficients we need for the Clausius trace. All of these are depending on different values of $C_{B}$, $C_{D}$, $d$, $m=M_{D}/M_{B}$ and the computation of them, where it was not possible in an analithical way, was performed numerically by the software Derive for Windows 6.0. In the Tab(\ref{tab:app1},\ref{tab:app2}, \ref{tab:app3}) are listed the values of all the coefficents for different parameters. Here we present how they were calulated, for example in the case $d=0.5$:  
\begin{eqnarray}
\begin{split}
	\left(\nu_{B}\right)_{M}&=3\int^{1}_{0}f_{B}\left(\xi_{B}\right)\xi^{2}_{B}d\xi_{B}\\
&=\frac{3}{H}\int^{1}_{0}\frac{2}{1+\left(C_{B}\xi_{B}\right)^{2}}\cdot\left\{\left[\frac{1+C^{2}_{B}}{1+(C_{B}\xi_{B})^{2}}\right]^{\frac{1}{2}}\cdot\cos^{-1}\left[\left(\frac{1+(C_{B}\xi_{B})^{2}}{1+C^{2}_{B}}\right)^{\frac{1}{2}}\right]-\left[\frac{C^{2}_{B}-(C_{B}\xi_{B})^{2}}{1+C^{2}_{B}}\right]^{\frac{1}{2}}\right\}\cdot\xi^{2}_{B}d\xi_{B}
\end{split}
\end{eqnarray}

\begin{eqnarray}
\begin{split}	
F_{B}\left(\xi_{B}\right)&=2\int^{1}_{\xi_{B}}f_{B}\left(\xi_{B}\right)\xi_{B}d\xi_{B}
=\frac{2}{H}\int^{1}_{\xi_{B}}\frac{2}{1+\left(C_{B}\xi_{B}\right)^{2}}\\
&\cdot\left\{\left[\frac{1+C^{2}_{B}}{1+(C_{B}\xi_{B})^{2}}\right]^{\frac{1}{2}}\cdot\cos^{-1}\left[\left(\frac{1+(C_{B}\xi_{B})^{2}}{1+C^{2}_{B}}\right)^{\frac{1}{2}}\right]-\left[\frac{C^{2}_{B}-(C_{B}\xi_{B})^{2}}{1+C^{2}_{B}}\right]^{\frac{1}{2}}\right\}\xi_{B} d\xi_{B} \\
&=\frac{4}{H}\left[Z\left(\xi_{B}\right)\right]^{1}_{\xi_{B}}
\end{split}
\end{eqnarray}
where:
\begin{eqnarray}
\begin{split}
Z\left(\xi_{B}\right)=&-\frac{1}{C^{2}_{B}}\sqrt{\frac{C^{2}_{B}+1}{(C_{B}\xi_{B})^{2}+1}}\cdot\cos^{-1}\left(\sqrt{\frac{1+(C_{B}\xi_{B})^{2}}{1+C^{2}_{B}}}\right)-\frac{1}{C^{2}_{B}}\ln\left[(C_{B}\xi_{B})^{2}+1\right]+\\
&+\frac{2}{C^{2}_{B}}\ln\left[C_{B}\sqrt{C^{2}_{B}+1}\cdot\sqrt{1-\xi^{2}_{B}}+C^{2}_{B}+1\right]-\frac{1}{C_{B}}\sqrt{1-\xi^{2}_{B}}\\
\end{split}
\end{eqnarray}

\begin{eqnarray}
	\nu_{\Omega B}=\frac{9}{16}\frac{1}{\left(\nu_{B}\right)^{2}_{M}}\int^{1}_{0}F^{2}_{B}\left(\xi_{B}\right)d\xi_{B}
\end{eqnarray}

\begin{eqnarray}
\begin{split}
	F_{D}\left(\xi_{D}\right)&=2\int^{1}_{\xi_{D}}f_{D}\left(\xi_{D}\right)\xi_{D} d\xi_{D}\\
&=2\int^{1}_{\xi_{D}}\frac{2}{1+\left(C_{D}\xi_{D}\right)^{0.5}}\cdot\xi_{D} d\xi_{D}\\
&=2\left[- \frac{4\ln(\sqrt{C_{D}\xi_{D}}+1)}{C^{2}_{D}} + \frac{4(C_{D}\xi_{D} + 3)\sqrt{C_{D}\xi_{D}}}{3C^{2}_{D}} - \frac{2\xi_{D}}{C_{D}}\right]^{1}_{\xi_{D}}\\
&=\frac{8\ln(\sqrt{C_{D}\xi_{D}}+1)}{C^{2}_{D}}- \frac{8\ln(\sqrt{C_{D}}+1)}{C^{2}_{D}}+\\
&\ \ \ \ -\frac{4\left[2(C_{D}\xi_{D}+3)\sqrt{C_{D}\xi_{D}}-\sqrt{C_{D}}(3\sqrt{C_{D}\xi_{D}}+2C_{D}-3\sqrt{C_{D}}+6)\right]}{3C^{2}_{D}}
\end{split}
\end{eqnarray}

\begin{eqnarray}
\begin{split}
	\left(\nu_{D}\right)_{M}&=3\int^{1}_{0}f_{D}\left(\xi_{D}\right)\xi^{2}_{D}d\xi_{D}\\
&=3\int^{1}_{0}\frac{2}{1+\left(C_{D}\xi_{D}\right)^{0.5}}\xi^{2}_{D}d\xi_{D}\\
&=3\left[- \frac{4\ln(\sqrt{C_{D}\xi_{D}}+1)}{C^3_{D}}+\frac{4(3C^{2}_{D}\xi^{2}_{D}+5C_{D}\xi_{D}+15)\sqrt{C_{D}\xi_{D}}}{15C^3_{D}}-\frac{\xi^{2}_{D}}{C_{D}}-\frac{2\xi_{D}}{C^{2}_{D}}\right]^{1}_{0}\\
&=\frac{12C^{2}_{D}-15C^{3/2}_{D}+20C_{D}-30C^{1/2}_{D}+60}{5C^{5/2}_{D}}-\frac{12\ln\sqrt{C_{D}+1}}{C^{3}_{D}}
\end{split}
\end{eqnarray}

\begin{eqnarray}
	\frac{dF_{D}\left(\xi_{D}\right)}{d\xi_{D}}=-\frac{4\xi_{D}}{1+\sqrt{C_{D}\xi_{D}}}
\end{eqnarray}
If we define $x=a_{B}/a_{D}$ we can exprime $\xi_{D}$ in terms of $\xi_{B}$ in the following way: $\xi_{D}=\xi_{B}\left(\frac{a_{B}}{a_{D}}\right)=\xi_{B}x$.
\begin{eqnarray}
w_{ext}\left(x\right)=\int^{x}_{0}F_{B}\left(\xi_{D}\right)\frac{dF_{D}\left(\xi_{D}\right)}{d\xi_{D}}\cdot\xi_{D}{d\xi_{D}}
\end{eqnarray}
\begin{eqnarray}
	\nu_{V}=-\frac{9}{8}\cdot\frac{1}{\left(\nu_{B}\right)_{M}\left(\nu_{D}\right)_{M}}\cdot m\cdot w_{ext}\left(x\right)
\end{eqnarray}
At the end, Clausius Virial, normalized by the factor $GM_{B}F/a_{D}$ can be expressed by:
\begin{eqnarray}
	\widetilde{V}_{B}\left(x\right)=-\frac{\nu_{\Omega B}}{x}-\frac{\nu_{V}}{x}
\end{eqnarray} 

\begin{table}
			\centering
			\begin{tabular}{|c|c|c|c|c|}
			\hline
			\hline
			$\log C_{B}$ & $C_{B}$ & $H$ & $\left(\nu_{B}\right)_{M}$ & $\nu_{\Omega B}$ \\
			\hline
0.00	&	1	&	0.0000	&		&		\\
1.00	&	10	&	9.1692	&	0.010343748	&	0.9039	\\
1.30	&	20	&	20.2438	&	0.001859053	&	1.2646	\\
1.48	&	30	&	31.3409	&	0.000657748	&	1.5575	\\
1.60	&	40	&	42.4431	&	0.000310761	&	1.8141	\\
1.70	&	50	&	53.5474	&	0.000172634	&	2.0472	\\
1.78	&	60	&	64.6527	&	0.000106399	&	2.2633	\\
1.85	&	70	&	75.7585	&	0.000070502	&	2.4666	\\
1.90	&	80	&	86.8647	&	0.000049279	&	2.6596	\\
1.95	&	90	&	97.9711	&	0.000035887	&	2.8441	\\
2.00	&	100	&	109.0777	&	0.000027000	&	3.0216	\\
\hline
\hline
	\end{tabular}
	\caption{\footnotesize Physical parameters of King's models as function of concentration, $C_{B}$.}
	\label{tab:app1}
\end{table}
\vspace{4cm}

\begin{table}
			\centering
			\begin{tabular}{|c|c|c|c|c|c|}
			\hline
			\hline
			$C_{D}$ & $\left(\nu_{D}\right)_{M}$ $\left[d=0.5\right]$ & $\left(\nu_{D}\right)_{M}$ $\left[d=1.0\right]$ & $\left(\nu_{D}\right)_{M}$ $\left[d=1.5\right]$ & $\left(\nu_{D}\right)_{M}$ $\left[d=2.0\right]$ & $\left(\nu_{D}\right)_{M}$ $\left[d=3.0\right]$ \\
			\hline
1	&	1.08223		&	1.15888		&	1.22741		&	1.28761				&	1.38629	\\
10	&	0.54627		&	0.25439		&	0.11255		&	0.05117				&	0.01382	\\
20	&	0.42054		&	0.13728		&	0.04247		&	0.01386				&	0.00225	\\
30	&	0.35746		&	0.09410		&	0.02359		&	0.00633				&	0.00076	\\
40	&	0.31735		&	0.07160		&	0.01547		&	0.00361				&	0.00035	\\
50	&	0.28880		&	0.05779		&	0.01113		&	0.00233				&	0.00019	\\
60	&	0.26709		&	0.04845		&	0.00849		&	0.00162			&	0.00011	\\
70	&	0.24982		&	0.04171		&	0.00676		&	0.00120				&	0.00007	\\
80	&	0.23564		&	0.03661		&	0.00554		&	0.00092			&	0.00005	\\
90	&	0.22371		&	0.03263		&	0.00465		&	0.00073			&	0.00004	\\
100	&	0.21349		&	0.02943		&	0.00397		&	0.00059			&	0.00003	\\
\hline
\hline
	\end{tabular}
	\caption{\footnotesize Physical parameters of the DM halo described by a cored power law density profile with exponent $d$.}
	\label{tab:app2}
\end{table}
\vspace{3cm}
\begin{table}
			\centering
			\begin{tabular}{|c|c|c|c|c|c|}
			\hline
			\hline
			$C_{u}$ & $\nu_{\Omega B}$ $\left[d=0.5\right]$ & $\nu_{\Omega B}$ $\left[d=1.0\right]$ & $\nu_{\Omega B}$ $\left[d=1.5\right]$ & $\nu_{\Omega B}$ $\left[d=2.0\right]$  & $\nu_{\Omega B}$ $\left[d=3.0\right]$ \\
			\hline
1	&	0.30470		&	0.30845		&	0.31111		&	0.31286				&	0.31457	\\
10	&	0.30812		&	0.32425		&	0.35305		&	0.40509			&	0.65953	\\
20	&	0.30903		&	0.32773		&	0.36251		&	0.43371			&	0.92480	\\
30	&	0.30950		&	0.32924		&	0.36639		&	0.44761			&	1.14502	\\
40	&	0.30981		&	0.33009		&	0.36850		&	0.45611			&	1.34119	\\
50	&	0.31003		&	0.33065		&	imp		&	0.46194				&	1.52170	\\
60	&	0.31021		&	0.33104		&	imp		&	0.46623				&	1.69090	\\
70	&	0.31034		&	0.33132		&	imp		&	0.46954				&	1.85141	\\
80	&	0.31046		&	0.33155		&	imp		&	0.47218				&	2.00495	\\
90	&	0.31056		&	0.33173		&	imp		&	0.47435					&	2.15274	\\
100	&	0.31064		&	0.33187		&	imp		&	0.47617				&	2.29566	\\
\hline
\hline
	\end{tabular}
	\caption{\footnotesize Physical parameters of the DM halo described by a cored power law density profile with exponent $d$.}
	\label{tab:app3}
\end{table}

\end{document}